\documentclass[runningheads,fleqn]{lncse}
\usepackage{multicol} 
\usepackage{makeidx}  
\usepackage{amssymb}
\usepackage{amsmath}
\usepackage{graphicx}

\begin{document}

\mainmatter

\title{Planar Solidification from Undercooled Melt: An Approximation
  of a Dilute Binary Alloy for a Phase-field Model}

\titlerunning{Planar solidification from undercooled melt}

\author{Denis Danilov}
\authorrunning{D. Danilov}

\institute{Laboratory of Condensed Matter Physics, Physical Faculty,
  Udmurt State University, 1 Universitetskaya Street, Izhevsk 426034,
  Russia\\ Email: ddanilov@uni.udm.ru}


\setcounter{page}{150}

\maketitle
\index{Danilov@Denis Danilov}
\begin{abstract}
  Planar solidification from an undercooled melt has been considered
  using the phase-field model. The solute and the phase fields have
  been found in the limit of small impurity concentration. These
  solutions in the limit of vanishing velocity of the interface motion
  give the equilibrium partition coefficient and the liquidus slope.
  Asymptotic expansions for the solute and for the phase fields, and
  the relation between the diffusive speed and the parameters of the
  phase field model have been found at high growth velocity. A
  comparison with numerical calculations is presented.
\end{abstract}

\section{Introduction}

Classical macroscopic models of solidification are based on the
assumption that the interface between regions of different phases have
zero thickness \cite{Kurz-Fisher}. Sharp-interface descriptions
require the introduction of separately derived nonequilibrium models
for the behavior of the interfacial temperature and of the solute
concentrations.  In contrast to the sharp-interface models, the
phase-field models
\cite{Wheeler-Boettinger-McFadden-1992,Bi-Sekerka-1998} describe the
bulk phases as well as the interface, i.e. these models treat the
system as a whole and eliminate the need to specify the interfacial
conditions separately. In phase-fields models, equilibrium behavior is
recovered at low growth velocities and nonequilibrium effects
naturally emerge at high growth velocities.  From a computational
point of view, the advantage of the phase-field formulation is that
the interface is not tracked but is given implicitly by the value of
the phase-field variable.  However, theoretical analysis solutions
were obtained by matching of separately derived ``inner'' (near the
interface) and ``outer'' (far from the interface) solutions in an
intermediate region
\cite{Wheeler-Boettinger-McFadden-1993,Elder-Grant-Provatas-Kosterlitz-2001}.
In the work \cite{Ahmad-Wheeler-Boettinger-McFadden-1998},
phase-fields and solute profiles have been obtained without using the
multiply-variable expansion, which leads to ``inner'' and ``outer''
solutions, but the profiles were obtained from a simplified set of
equations in which the interface kinetics was eliminated.

In this work an analytical solution of the phase-field model for
planar solidification from an undercooled melt as an expansion in
terms of the solute concentration far from the interface is presented.
This solution is valid for the bulk phases as well as for the
interface and includes the interface kinetics. To verify the expansion,
numerical simulations of the problem were carried out.

\section{The Model}

Let us consider the motion of a planar solid-liquid interface during
the solidification of a binary alloy. The governing equations for the
phase-field $\phi(x,t)$ and for the solute concentration $c(x,t)$ are
given by the following set of equations
\cite{Wheeler-Boettinger-McFadden-1992}
\begin{equation}
  \label{eq:phi-concentration-general}
    \frac{\partial \phi}{\partial t}
     = M_1 \left(
      \epsilon^2 \frac{\partial^2\phi}{\partial x^2}
      - \frac{\partial f}{\partial \phi}
    \right),
    \qquad
    \frac{\partial c}{\partial t}
     = \frac{\partial}{\partial x}
    M_2 c(1-c) \frac{\partial}{\partial x}
    \frac{\partial f}{\partial c},
\end{equation}
which satisfy the conservation of solute and which have been derived
under the assumption that the total free-energy decreases
monotonically in time. The parameters $M_1$, $M_2$, and $\epsilon$ are
related to the growth kinetics, to the diffusion coefficient, and to
the surface energy respectively.

In the approach of ideal solution, the free-energy density
$f(\phi,c,T)$ at a temperature $T$ is given by
\cite{Wheeler-Boettinger-McFadden-1992}
\begin{equation}
  \label{eq:f(phi,c,T)}
  f(\phi,c,T)
  = c f_B(\phi,T)
  + (1-c) f_A(\phi,T)
  + \frac{RT}{v_m}[c \ln c + (1-c)\ln(1-c)],
\end{equation}
where $R$ is the universal gas constant, $v_m$ is the molar volume,
which is assumed to be constant. Free-energy densities of the pure
materials $A$ and $B$ are assumed to be of the form
\cite{Wheeler-Boettinger-McFadden-1992}
\begin{equation}
  \label{eq:f_i(phi,T)}
  f_i(\phi,T)
  = W_i \int_{0}^{\phi} p(p-1)
  \left(
    p-\frac{1}{2}-\beta_i(T)
  \right) dp
  = \frac{W_i}{4} g(\phi)
  + \frac{W_i \beta_i(T)}{6} p(\phi)
\end{equation}
where
\begin{equation}
  \label{eq:g_and_p}
  g(\phi) = \phi^4-2\phi^3+\phi^2,
  \quad
  p(\phi) = \phi^2(3-2\phi),
  \quad
  \frac{W_i\beta_i(T)}{6}
  = L_i \frac{T-T_i}{T_i}.
\end{equation}
Here $W_i$ is a constant, $L_i$ is the latent heat per unit volume,
$T_i$ is the melting point of material $i=A,B$.

To investigate steady-state interface motion, the boundary conditions
will be taken in the form
\begin{equation}
  \label{eq:far-field-conditions}
  c|_{x\rightarrow\infty} = c_{\infty},
    \quad
    \phi|_{x\rightarrow\infty} = 0,
    \qquad
    \frac{\partial c}{\partial x}|_{x\rightarrow -\infty} = 0,
    \quad
    \phi|_{x\rightarrow -\infty} = 1,
\end{equation}
and the moving frame $z$ will be adopted according to the equation
\begin{equation}
  \label{eq:moving-frame}
  z = x - Vt,
\end{equation}
so that the interface (given by $\phi=1/2$) corresponds to $z=0$ at
this frame.  In the moving frame, equations
(\ref{eq:phi-concentration-general}) may be written by using
(\ref{eq:f(phi,c,T)}) as
\begin{equation}
  \label{eq:phi-moving-frame}
  \epsilon^2 \frac{\partial^2 \phi}{\partial z^2}
  +  \frac{V}{M_1} \frac{\partial \phi}{\partial z}
  - \left[
    \frac{\partial f_A}{\partial \phi}
    + c \left(
      \frac{\partial f_B}{\partial \phi}
      - \frac{\partial f_A}{\partial \phi}
    \right)
  \right] = 0,
\end{equation}
\begin{equation}
  \label{eq:concentration-moving-frame}
  \frac{M_2 R T}{v_m} \frac{\partial^2 c}{\partial z^2}
  + V \frac{\partial c}{\partial z}
  + M_2 \frac{\partial}{\partial z}
  \left[
    c(1-c)
    \left(
      \frac{\partial f_B}{\partial \phi}
      - \frac{\partial f_A}{\partial \phi}
    \right)
    \frac{\partial \phi}{\partial z}
  \right] = 0.
\end{equation}
For further analysis we introduce the notation for the diffusion
coefficient $D=M_2 R T/v_m$ and integrate equation
(\ref{eq:concentration-moving-frame}) with respect to $z$ over an
interval $[z,\infty)$. Employing the boundary conditions
(\ref{eq:far-field-conditions}) gives
\begin{equation}
  \label{eq:concentration-moving-frame-int}
  D \frac{\partial c}{\partial z}
  + V (c - c_{\infty})
  + M_2 c(1-c)
  \left(
    \frac{\partial f_B}{\partial \phi}
    - \frac{\partial f_A}{\partial \phi}
  \right)
  \frac{\partial \phi}{\partial z}
  = 0.
\end{equation}

\section{An Approximation of a Small Impurity Concentration}

Let us examine a dilute binary alloy with initial impurity
concentration $c_{\infty}\ll 1$.  For this purpose we shall consider
the expansion in terms $c_{\infty}$ for the phase-field, the solute
concentration, and the temperature
\begin{align}
  & \phi(z) = \phi_0(z) + c_{\infty} \phi_1(z) + O(c_{\infty}^2),
  \label{eq:phi-expansion-dilute} \\
  & c(z) = c_{\infty} c_1(z) + O(c_{\infty}^2),
  \label{eq:concentration-expansion-dilute} \\
  & T =  T_0 + c_{\infty} T_1 + O(c_{\infty}^2),
  \label{eq:T-expansion-dilute}
\end{align}

As a result, in the zeroth-order approximation, from
(\ref{eq:phi-moving-frame}) by using (\ref{eq:f_i(phi,T)}) we obtain
the equation for $\phi_0(z)$
\begin{equation}
  \label{eq:phi-0}
  \epsilon^2 \frac{\partial^2 \phi_0}{\partial z^2}
  +  \frac{V}{M_1} \frac{\partial \phi_0}{\partial z}
  - \left(
    \frac{W_A}{4} \left.\frac{\partial g}{\partial
        \phi}\right|_{\phi_0}
    + \frac{W_A\beta_A}{6} \left.\frac{\partial p}{\partial
        \phi}\right|_{\phi_0}
  \right) = 0,
\end{equation}
The solution of (\ref{eq:phi-0}), which satisfies the boundary
conditions (\ref{eq:far-field-conditions}), describes the phase field
during the solidification of a pure material $A$ and has the form
\begin{equation}
  \label{eq:phi-0-solution}
  \phi_0(z) = \frac{1}{2}
  \left[
    1 - \tanh\left(\frac{z}{2l_A}\right)
  \right],
\end{equation}
where $l_A$ is the interface thickness and the interface velocity
$V$ is related to the temperature $T_0$ by the relationship
\begin{equation}
  \label{eq:V-T}
  V = - M_1 l_A W_A \beta_A(T_0)
  = \frac{6 M_1 l_A L_A}{T_A} (T_A - T_0),
\end{equation}
which determines the interface kinetics for a pure material. The
relationship (\ref{eq:V-T}) corresponds to the model of collision
limited growth \cite{Turnbull-1981} with a kinetic coefficient
\begin{equation}
  \label{eq:mu}
  \mu = \frac{6 M_1 l_A L_A}{T_A},
  \quad
  T_0 = T_A - \frac{V}{\mu}.
\end{equation}

In the first-order approximation, the equations
(\ref{eq:phi-moving-frame}) and
(\ref{eq:concentration-moving-frame-int}) give
\begin{multline}
  \label{eq:phi-1}
  \epsilon^2 \frac{\partial^2 \phi_1}{\partial z^2}
  + \frac{V}{M_1} \frac{\partial \phi_1}{\partial z}
  - \Bigg[
  \phi_1 \left.
    \frac{\partial^2 f_A}{\partial \phi^2}
  \right|_{\phi_0, T_0}
  + T_1 \left.
    \frac{\partial^2 f_A}{\partial T \partial \phi}
  \right|_{\phi_0, T_0} \\
  + c_1 \left.
    \left(
      \frac{\partial f_B}{\partial \phi}
      - \frac{\partial f_A}{\partial \phi}
    \right)
  \right|_{\phi_0, T_0}
  \Bigg]
  = 0,
\end{multline}
\begin{equation}
  \label{eq:concentration-1-dF}
  \frac{\partial c_1}{\partial z}
  + \frac{V}{D} (c_1 - 1)
  + c_1 \frac{\partial \Delta F}{\partial z}
  = 0,
\end{equation}
where function $\Delta F(z)$ is defined by expression
\begin{equation}
  \label{eq:dF(z)}
  \Delta F(z) = \frac{v_m}{RT_0}
  \left[
    f_B(\phi_0(z), T_0) - f_A(\phi_0(z), T_0)
  \right],
\end{equation}
Taking into account the boundary conditions
(\ref{eq:far-field-conditions}), the solution of
(\ref{eq:concentration-1-dF}) is
\begin{equation}
  \label{eq:concentration-1-dF-solution-final}
  c_1(z) = \frac{V}{D}
  \int_{-\infty}^{z}
  e^{ V(z'-z)/D + \Delta F(z') - \Delta F(z) }
  dz'
\end{equation}

The point $z=0$ corresponds to the solid-liquid interface, i.~e.
$\phi(0)=\phi_0(0)=1/2$, and to the inflection of the phase-field,
therefore
\begin{equation}
  \label{eq:phi1-z=0}
  \phi_1|_{z=0} = 0,
  \quad
  \left.\frac{\partial^2 \phi_1}{\partial z^2}\right|_{z=0} = 0.
\end{equation}
As for metallic systems $M_1\sim 10^{8}$\,cm$^3$/J$\cdot$s
\cite{Wheeler-Boettinger-McFadden-1992}, then the relation $V/M_1\ll
1$ will be satisfied. Taking into account the conditions
(\ref{eq:phi1-z=0}) at $z=0$, one can find from (\ref{eq:phi-1}) the
relationship between $T_1$ and $c_1$
\begin{equation}
  \label{eq:T_1-c_1-z=0}
  T_1 =
  - \frac{\displaystyle\left.
      \left(
        \frac{\partial f_B}{\partial \phi}
        - \frac{\partial f_A}{\partial \phi}
      \right)
    \right|_{\phi_0, T_0}}%
  {\displaystyle\left.
      \frac{\partial^2 f_A}{\partial T \partial \phi}
    \right|_{\phi_0, T_0}}
  \, c_1|_{z=0}.
\end{equation}
Equations (\ref{eq:concentration-expansion-dilute}),
(\ref{eq:T-expansion-dilute}), (\ref{eq:mu}), and
(\ref{eq:T_1-c_1-z=0}) give the expression for the temperature
\begin{equation}
  \label{eq:T}
  T = T_A - \frac{V}{\mu} - m(V) c|_{z=0},
\end{equation}
where the function $m(V)$ is defined by
\begin{equation}
  \label{eq:m(V)}
    m(V) =
    \frac{\displaystyle\left.
        \left(
          \frac{\partial f_B}{\partial \phi}
          - \frac{\partial f_A}{\partial \phi}
        \right)
      \right|_{\phi_0, T_0}}%
    {\displaystyle\left.
        \frac{\partial^2 f_A}{\partial T \partial \phi}
      \right|_{\phi_0, T_0}}
    = \left.
      \frac{W_B\beta_B(T)-W_A\beta_A(T)}%
      {W_A \displaystyle\frac{\partial \beta_A}{\partial T}}
    \right|_{T=T_A-V/\mu}
\end{equation}

\section{Stationary Interface}

Now we consider the solute field due to the stationary solid-liquid
interface, at $V=0$, which corresponds to the equilibrium state. Let
us introduce the definitions for the bulk concentrations in the liquid
$c_L=c(z\rightarrow\infty)$ and in the solid
$c_S=c(z\rightarrow-\infty)$.  For a stationary interface, first it
follows from (\ref{eq:mu}) that $T_0=T_A$, and second the equation
(\ref{eq:concentration-1-dF}) gives the solute field
\begin{equation}
  \label{eq:stationary-front-c(z)}
  \frac{c(z)}{c_L} = e^{-\Delta F(z)},
\end{equation}
which leads to the equilibrium partition coefficient
\begin{equation}
  \label{eq:stationary-front-k_e}
  k_e = \frac{c_S}{c_L}
  = \exp[-\Delta F(z\rightarrow-\infty)]
  = \exp\left(
    - \frac{v_m L_B (T_A-T_B)}{R T_A T_B}
  \right).
\end{equation}
This expression has the same form as in
\cite{Ahmad-Wheeler-Boettinger-McFadden-1998}, where the equilibrium
partition coefficient has been obtained from a simplified set of
equations in which the effects of interface attachment kinetics are
eliminated.

Using (\ref{eq:g_and_p}), from equation (\ref{eq:m(V)}) at $T_0=T_A$
we obtain
\begin{equation}
  \label{eq:m_0}
  m_0 = m(0)
  = \frac{L_B}{L_A}
  \frac{T_A}{T_B}
  (T_A - T_B).
\end{equation}
The solute concentration at the interface $z=0$, in accordance with
(\ref{eq:stationary-front-c(z)}), is
\begin{equation}
  \label{eq:stationary-front-c(0)}
  c|_{z=0}
  = c_L \exp \Bigg(
    - \frac{v_m}{R T_A}
    \Bigg[
      \frac{W_B-W_A}{64}
      + \frac{1}{2} \frac{L_B(T_A-T_B)}{T_B}
    \Bigg]
  \Bigg).
\end{equation}
Thus equations (\ref{eq:T}), (\ref{eq:m_0}), and
(\ref{eq:stationary-front-c(0)}) give the temperature at the
equilibrium solid-liquid interface
\begin{multline}
  \label{eq:stationary-front-T}
  T = T_A - m_0 c|_{z=0}
  =  T_A - m_0 \exp \Bigg(
    - \frac{v_m}{R T_A}
    \Bigg[
      \frac{W_B-W_A}{64} \\
      + \frac{1}{2} \frac{L_B(T_A-T_B)}{T_B}
    \Bigg]
  \Bigg) c_L
  = T_A - m_e c_L,
\end{multline}
which yields the following equation for equilibrium liquidus slope
\begin{equation}
  \label{eq:stationary-front-m_e}
  m_e  = \frac{L_B}{L_A}\frac{T_A}{T_B}(T_A - T_B)
  \exp \Bigg(
    - \frac{v_m}{R T_A}
    \Bigg[
       \frac{W_B-W_A}{64}
       + \frac{L_B(T_A-T_B)}{2 T_B}
    \Bigg]
  \Bigg).
\end{equation}

\section{Large-Velocity Asymptotics}

To investigate the behavior of the solute concentration under rapid
solidification conditions, we consider an asymptotic analysis in the
limit $V\gg 1$. The solution
(\ref{eq:concentration-1-dF-solution-final}) can be written in the
form
\begin{equation}
  \label{eq:concentration-1-dF-solution-large-V}
  c_1(z) =
  \frac{V}{D}
  \int_0^{\infty}
  e^{ -Vz'/D + \Delta F(z-z') -\Delta F(z) }
  d z'.
\end{equation}
Applying the Laplace's theorem about asymptotic expansions
\cite{Olver} to this integral, we obtain the solute field with an
accuracy to second order
\begin{equation}
  \label{eq:concentration-1-dF-solution-large-V-asympt}
  c_1(z) = \frac{c(z)}{c_{\infty}} =
  1 - \frac{D}{V} \frac{\partial \Delta F}{\partial z}.
\end{equation}
According to (\ref{eq:concentration-1-dF-solution-large-V-asympt}),
a maximum value of solute is reached at $z=0$. Therefore the segregation
coefficient at large velocities is given by
\begin{multline}
  \label{eq:large-V-k}
  k = \frac{c_{\infty}}{c|_{z=0}}
  = \frac{1}{1-\displaystyle\frac{D}{V}
    \left.
      \frac{\partial \Delta F}{\partial z}
    \right|_{z=0}}
  = 1 + \frac{D}{V}
  \left.
    \frac{\partial \Delta F}{\partial z}
  \right|_{z=0} \\
  = 1 - \frac{D}{4 l_A V} \frac{v_m}{RT}
  \left.
    \left(
      \frac{\partial f_B}{\partial \phi}
      - \frac{\partial f_A}{\partial \phi}
    \right)
  \right|_{\phi=1/2}.
\end{multline}

For the sharp-interface model the nonequilibrium segregation
coefficient has been derived by Aziz \cite{Aziz-1982} and in the limit
of dilute alloy is
\begin{equation}
  \label{eq:k-Aziz}
  k^{(a)} = \frac{k_e + V/V_D}{1 + V/V_D},
\end{equation}
where diffusive speed $V_D$ is a characteristic trapping velocity.
For $V/V_D\gg 1$ the segregation coefficient $k^{(a)}$ can be
approximated by
\begin{equation}
  \label{eq:k-Aziz-large-V}
  k^{(a)} = 1 - (1-k_e)\frac{V_D}{V}.
\end{equation}
A comparison of the equations (\ref{eq:large-V-k}) and
(\ref{eq:k-Aziz-large-V}) gives
\begin{equation}
  \label{eq:large-V-VD}
  V_D = \frac{D}{4 l_A (1-k_e)} \frac{v_m}{RT}
  \left.
    \left(
      \frac{\partial f_B}{\partial \phi}
      - \frac{\partial f_A}{\partial \phi}
    \right)
  \right|_{\phi=1/2, T_0}
\end{equation}
According to (\ref{eq:large-V-VD}), the diffusive speed $V_D$ depends
on the diffusion coefficient, on the temperature, and on the
difference of free energy densities of the pure materials.

\section{Numerical Calculations}

To validate the obtained solution, we will compare it with numerical
solutions of the steady-state governing equations
(\ref{eq:phi-moving-frame}) and (\ref{eq:concentration-moving-frame}).
The material parameters used in the numerical calculations are given
in Table~1. These parameters are similar to parameters employed in
\cite{Ahmad-Wheeler-Boettinger-McFadden-1998} and correspond to the
alloy Ni--Cu \cite{Wheeler-Boettinger-McFadden-1992}.

\begin{table}[t]
  \begin{center}
    \caption{Thermophysical properties used in calculations}
    \label{tab:therm_phys_param}
    \begin{tabular}{cccc}
      \hline
      \hline
      Parameter & Value &  & Reference \\
      \hline
      $L_A$ & $2350$ & J/cm$^{2}$ &
      \cite{Ahmad-Wheeler-Boettinger-McFadden-1998} \\
      $L_B$ & $1725$ & J/cm$^{2}$ &
      \cite{Ahmad-Wheeler-Boettinger-McFadden-1998} \\
      $T_A$ & $1728$ & K &
      \cite{Ahmad-Wheeler-Boettinger-McFadden-1998} \\
      $T_B$ & $1358$ & K &
      \cite{Ahmad-Wheeler-Boettinger-McFadden-1998} \\
      $D$ & $10^{-5}$ & cm$^{2}$/ñ &
      \cite{Ahmad-Wheeler-Boettinger-McFadden-1998} \\
      $\sigma$ & $2.8 \times 10^{-5}$ & J/cm$^{2}$ &
      \cite{Ahmad-Wheeler-Boettinger-McFadden-1998} \\
      $l_A$ & $6.48 \times 10^{-8}$ & cm &
      \cite{Ahmad-Wheeler-Boettinger-McFadden-1998} \\
      $v_m$ & $7.4$ & cm$^3$/mol &
      \cite{Wheeler-Boettinger-McFadden-1992} \\
      $\mu$ & $24$ & cm/(s$\cdot$K) &
      \cite{Galenko-Danilov-1999} \\
      \hline
    \end{tabular}
  \end{center}
\end{table}

The value of kinetic coefficient was set to $\mu=24$~cm/(s$\cdot$K)
according to the work \cite{Galenko-Danilov-1999} in which good
agreements between the model for dendritic growth and experimental
data have been achieved.  The far-field concentration was set to
$c_{\infty}=0.1$.  The parameter $\epsilon$ is related to the surface
energy $\sigma$ and to the interface thickness $l_A$ by equation
$\epsilon^2=6\sigma l_A$
\cite{Wheeler-Boettinger-McFadden-1992,Ahmad-Wheeler-Boettinger-McFadden-1998}.
The parameters $W_A$ and $W_B$ were chosen to be equal and are given
as $W_A=W_B=W=12\sigma/l_A$
\cite{Ahmad-Wheeler-Boettinger-McFadden-1998}.

Employing (\ref{eq:f_i(phi,T)}) and (\ref{eq:g_and_p}), a
finite-difference approximation of the equations
(\ref{eq:phi-moving-frame}) and (\ref{eq:concentration-moving-frame})
on a uniform grid gives the following set of equations
\begin{multline}
  \label{eq:numeric-phi}
  \frac{\phi_i^{k+1}-\phi_i^{k}}{\Delta t}
  = \frac{V}{M_1 \epsilon}
  \frac{\phi_{i+1}^{k}-\phi_{i-1}^{k}}{2\Delta x}
  + \frac{\phi_{i+1}^{k}-2\phi_{i}^{k}+\phi_{i-1}^{k}}{\Delta x^2} \\
  - \frac{W}{4}\left(\frac{\partial g}{\partial \phi}\right)_i^k
  - \frac{W(c\beta_B+(1-c)\beta_A)}{6}
  \left(\frac{\partial p}{\partial \phi}\right)_i^k,
\end{multline}
\begin{multline}
  \label{eq:numeric-concentration}
  \frac{c_i^{k+1}-c_i^{k}}{\Delta t}
  = \frac{V}{M_1 \epsilon}
  \frac{c_{i+1}^{k}-c_{i-1}^{k}}{2\Delta x} \\
  + \frac{D}{M_1\epsilon^2}
  \frac{c_{i+1}^{k}-2c_{i}^{k}+c_{i-1}^{k}}{\Delta x^2}
  + \frac{M_2}{M_1\epsilon^2}
  \frac{{q}_{i+1}^{k}-{q}_{i-1}^{k}}{2\Delta x},
\end{multline}
\begin{equation}
  \label{eq:q}
  {q}_i^k
  = c_i^k(1-c_i^k)
    \left[
      \left(\frac{\partial f_B}{\partial \phi}\right)_i^k
      - \left(\frac{\partial f_A}{\partial \phi}\right)_i^k
    \right]
    \frac{\phi_{i+1}^{k}-\phi_{i-1}^{k}}{2\Delta x},
\end{equation}
which determine the discrete functions of the solute and of the phase
field at a point $z=i\Delta x$ on a $(k+1)$th step of iteration with a
time step $\Delta t$. To equalize the velocity $V$ of the moving frame
with the velocity of the interface, we introduce the equation
\begin{equation}
  \label{eq:V-relax}
  \tau_f \frac{d V(\tau)}{d\tau}
  = V_s(\tau)
  = M_1 \epsilon \frac{d z_f}{d \tau},
\end{equation}
where $V_s(\tau)$ is the shift velocity of the interface relative to
the moving frame, $\tau_f$ is an accessory parameter, $\tau$ is a time
variable corresponding to the iteration process, and the interface
coordinate $z_f$ is given by the condition $\phi(z_f)=1/2$. A
finite-difference approximation of equation (\ref{eq:V-relax}) gives
\begin{equation}
  \label{eq:numeric-V-relax}
  \frac{V^{k+1}-V^k}{\Delta t}
  = \frac{\epsilon M_1}{\tau_f}
  \frac{z_f^{k+1}-z_f^k}{\Delta t}.
\end{equation}

Results of numerical calculations of the equations
(\ref{eq:numeric-phi})--(\ref{eq:numeric-V-relax}) are shown in the
following figures. Fig.~1 shows a good agreement between the
analytical solution [equations
(\ref{eq:concentration-expansion-dilute}),
(\ref{eq:concentration-1-dF-solution-final}), (\ref{eq:T}), and
(\ref{eq:m(V)})] and the numerical results in the wide range of
undercoolings.

\begin{figure}[h]
  \begin{center}
    \includegraphics[width=82mm]{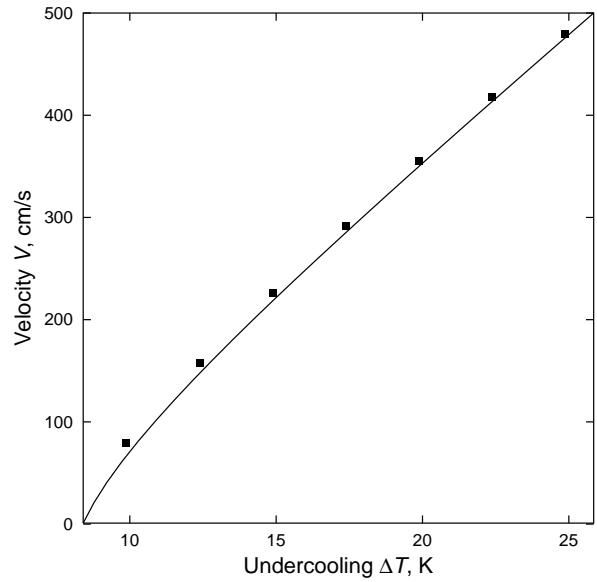}
    \caption{Kinetic curve ``interface velocity $V$~--- initial
      undercooling $\Delta T$''. The solid curve shows the
      approximated solution, data points denote the result of the
      numerical calculations.}
    \label{fig:V-dT}
  \end{center}
\end{figure}

Fig.~2 shows the solute and the phase fields. At the interface, the
solute field is smooth as against sharp-interface models, which gives
a jump in concentration.  At low undercoolings, the analytical
solution gives some excessive value of solute concentration in the
vicinities of the interface. The series of numerical calculations
given in Fig.~2 demonstrate the degeneration of the solute to uniform
field with increasing interface velocity and undercooling. Such
behavior of the solute field is in accordance with the equation
(\ref{eq:concentration-1-dF-solution-large-V-asympt}), in which the
deviation from the far-field concentration is proportional to $1/V$.

\begin{figure}[h]
  \begin{center}
    \includegraphics[width=82mm]{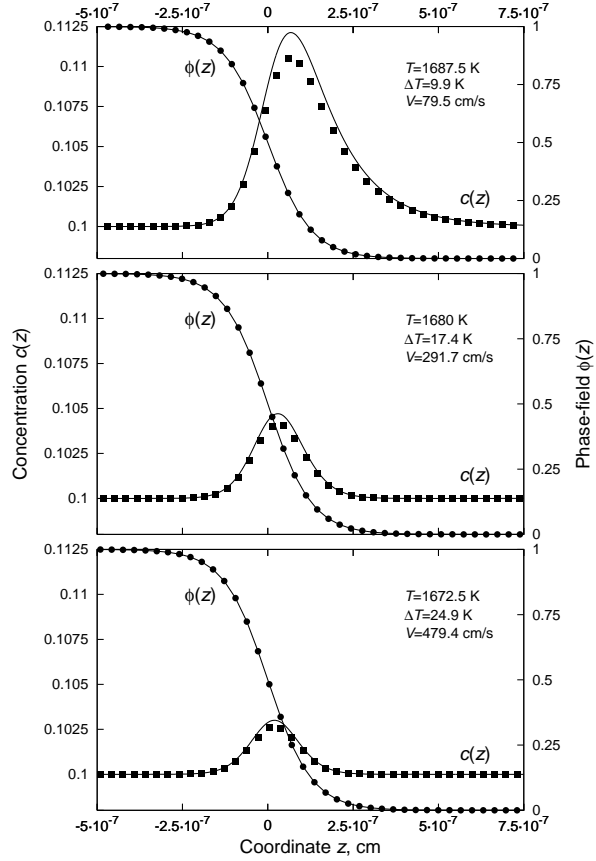}
    \caption{Solute profiles and the phase field for different values of
      the initial undercooling. The solid curve shows the approximated
      solution, data points denote the result of numerical
      calculations.}
    \label{fig:concentration-phi-z}
  \end{center}
\end{figure}

\section{Conclusions}

Planar solidification from undercooled melt was investigated by
numerical and analytical methods in the limit of small impurity
concentration.  Comparing the obtained results with those of
\cite{Ahmad-Wheeler-Boettinger-McFadden-1998} we note that while
equilibrium partition coefficient (\ref{eq:stationary-front-k_e}) has
the same form as in work
\cite{Ahmad-Wheeler-Boettinger-McFadden-1998}, the expression
(\ref{eq:stationary-front-m_e}) for the liquidus slope differs from
the one given in the work
\cite{Ahmad-Wheeler-Boettinger-McFadden-1998}. However the evaluated
equilibrium liquidus slope has very close values, $m_e=306.2$~K from
equation (\ref{eq:stationary-front-m_e}) and $m_e=306.9$~K from
equation (A13) in \cite{Ahmad-Wheeler-Boettinger-McFadden-1998}.  The
diffusive speed has a weak dependence on the temperature,
Eq.~(\ref{eq:large-V-VD}), and can be accepted to be constant as well
as the diffusive speed obtained in
\cite{Ahmad-Wheeler-Boettinger-McFadden-1998}. The numerical
calculations presented here show a good agreement with the analytical
solution that indicate a sufficient accuracy of the expansion
(\ref{eq:phi-expansion-dilute})--(\ref{eq:T-expansion-dilute}).

The obtained solutions simultaneously are valid for the bulk phases
and the interface. Therefore they can be suitable for the purpose of
numerical simulations: for instance to construct initial splitting of
adaptive finite-element grid and as a first approximation for the
numerical solution of transcendental equations.



\begin{thebibliography}{99}

\bibitem{Kurz-Fisher} Kurz\,W., Fisher\,D.\,J.: Fundamentals of
  Solidification, 3rd ed. Aedermannsdorf:\,Trans Tech Publication,
  1992. 305\,p.

\bibitem{Wheeler-Boettinger-McFadden-1992} Wheeler\,A.\,A.,
  Boettinger\,W.\,J., McFadden\,G.\,B.: Phase-field model for isothermal
  phase transitions in binary alloys. Phys. Rev.~A \textbf{45} (1992)
  7424--7438.

\bibitem{Bi-Sekerka-1998} Bi\,Z., Sekerka\,R.\,F.: Phase-field model of
  solidification of a binary alloy. Physica~A \textbf{261} (1998)
  95--106.

\bibitem{Wheeler-Boettinger-McFadden-1993} Wheeler\,A.\,A.,
  Boettinger\,W.\,J., McFadden\,G.\,B.: Phase-field model of solute
  trapping during solidification. Phys. Rev.~E \textbf{47} (1993)
  1893--1909.

\bibitem{Elder-Grant-Provatas-Kosterlitz-2001} Elder\,K.\,R.,
  Grant\,M., Provatas\,N., Kosterlitz\,J.\,M.: Sharp interface limits
  of phase-field models. Phys. Rev.~E \textbf{64} (2001) 021604.

\bibitem{Ahmad-Wheeler-Boettinger-McFadden-1998} Ahmad\,N.\,A.,
  Wheeler\,A.\,A., Boettinger\,W.\,J., McFadden\,G.\,B.: Solute
  trapping and solute drag in a phase-field model of rapid
  solidification. Phys. Rev.~E \textbf{58} (1998) 3436--3450.

\bibitem{Turnbull-1981} Turnbull\,D.: Metastable structures in
  metallurgy. Metall. Trans. A. \textbf{12} (1981) 695--708.

\bibitem{Olver} Olver\,F.\,W.\,J.: Asymptotics and Special Functions.
  New York:\,Academic Press, 1974. 584~p.

\bibitem{Aziz-1982} Aziz\,M.\,J.: Model for solute redistribution
  during rapid solidification. J.~Appl. Phys. \textbf{53} (1982)
  1158--1168.

\bibitem{Galenko-Danilov-1999} Galenko\,P.\,K., Danilov\,D.\,A.: Model
  for free dendritic alloy growth under interfacial and bulk phase
  nonequilibrium conditions. J.~Crystal Growth \textbf{197} (1999)
  992--1002.

\end{thebibliography}
\end{document}